\documentclass[sigconf, nonacm]{acmart}

\renewcommand\footnotetextcopyrightpermission[1]{} 
\acmConference{}{}{}
\acmBooktitle{}
\acmYear{}
\acmPrice{}
\acmDOI{}
\acmISBN{}
\usepackage{fancyhdr}
\AtBeginDocument{%
  }

\usepackage{booktabs}
\usepackage{fontawesome5}
\usepackage{subcaption}
\usepackage{listings}
\usepackage{makecell}
\usepackage{tabularx}
\usepackage{tikz}
\usepackage{url}
\usepackage{xcolor}
\usepackage{hyperref}
\usepackage{minted}

\usetikzlibrary{calc, positioning}
\usepackage{cleveref}

\definecolor{hgfblue}{RGB}{0, 90, 160}
\colorlet{hgfblue10}{hgfblue!10!white}
\colorlet{hgfblue20}{hgfblue!20!white}
\colorlet{hgfblue30}{hgfblue!30!white}
\colorlet{hgfblue40}{hgfblue!40!white}
\colorlet{hgfblue50}{hgfblue!50!white}
\colorlet{hgfblue60}{hgfblue!60!white}
\colorlet{hgfblue70}{hgfblue!70!white}
\colorlet{hgfblue80}{hgfblue!80!white}
\colorlet{hgfblue90}{hgfblue!90!white}

\definecolor{hgfdarkblue}{RGB}{0, 40, 100}
\colorlet{hgfdarkblue10}{hgfdarkblue!10!white}
\colorlet{hgfdarkblue20}{hgfdarkblue!20!white}
\colorlet{hgfdarkblue30}{hgfdarkblue!30!white}
\colorlet{hgfdarkblue40}{hgfdarkblue!40!white}
\colorlet{hgfdarkblue50}{hgfdarkblue!50!white}
\colorlet{hgfdarkblue60}{hgfdarkblue!60!white}
\colorlet{hgfdarkblue70}{hgfdarkblue!70!white}
\colorlet{hgfdarkblue80}{hgfdarkblue!80!white}
\colorlet{hgfdarkblue90}{hgfdarkblue!90!white}

\definecolor{hgflightblue}{RGB}{20, 200, 255}
\colorlet{hgflightblue10}{hgflightblue!10!white}
\colorlet{hgflightblue20}{hgflightblue!20!white}
\colorlet{hgflightblue30}{hgflightblue!30!white}
\colorlet{hgflightblue40}{hgflightblue!40!white}
\colorlet{hgflightblue50}{hgflightblue!50!white}
\colorlet{hgflightblue60}{hgflightblue!60!white}
\colorlet{hgflightblue70}{hgflightblue!70!white}
\colorlet{hgflightblue80}{hgflightblue!80!white}
\colorlet{hgflightblue90}{hgflightblue!90!white}

\definecolor{hgfgreen}{RGB}{140, 180, 35}
\colorlet{hgfgreen10}{hgfgreen!10!white}
\colorlet{hgfgreen20}{hgfgreen!20!white}
\colorlet{hgfgreen30}{hgfgreen!30!white}
\colorlet{hgfgreen40}{hgfgreen!40!white}
\colorlet{hgfgreen50}{hgfgreen!50!white}
\colorlet{hgfgreen60}{hgfgreen!60!white}
\colorlet{hgfgreen70}{hgfgreen!70!white}
\colorlet{hgfgreen80}{hgfgreen!80!white}
\colorlet{hgfgreen90}{hgfgreen!90!white}

\definecolor{hgfgray}{RGB}{90, 105, 110}
\colorlet{hgfgray10}{hgfgray!10!white}
\colorlet{hgfgray20}{hgfgray!20!white}
\colorlet{hgfgray30}{hgfgray!30!white}
\colorlet{hgfgray40}{hgfgray!40!white}
\colorlet{hgfgray50}{hgfgray!50!white}
\colorlet{hgfgray60}{hgfgray!60!white}
\colorlet{hgfgray70}{hgfgray!70!white}
\colorlet{hgfgray80}{hgfgray!80!white}
\colorlet{hgfgray90}{hgfgray!90!white}

\definecolor{hgfhighlight}{RGB}{205, 238, 251}
\colorlet{hgfhighlight10}{hgfhighlight!10!white}
\colorlet{hgfhighlight20}{hgfhighlight!20!white}
\colorlet{hgfhighlight30}{hgfhighlight!30!white}
\colorlet{hgfhighlight40}{hgfhighlight!40!white}
\colorlet{hgfhighlight50}{hgfhighlight!50!white}
\colorlet{hgfhighlight60}{hgfhighlight!60!white}
\colorlet{hgfhighlight70}{hgfhighlight!70!white}
\colorlet{hgfhighlight80}{hgfhighlight!80!white}
\colorlet{hgfhighlight90}{hgfhighlight!90!white}

\definecolor{hgfmint}{RGB}{5, 229, 186}
\colorlet{hgfmint10}{hgfmint!10!white}
\colorlet{hgfmint20}{hgfmint!20!white}
\colorlet{hgfmint30}{hgfmint!30!white}
\colorlet{hgfmint40}{hgfmint!40!white}
\colorlet{hgfmint50}{hgfmint!50!white}
\colorlet{hgfmint60}{hgfmint!60!white}
\colorlet{hgfmint70}{hgfmint!70!white}
\colorlet{hgfmint80}{hgfmint!80!white}
\colorlet{hgfmint90}{hgfmint!90!white}

\definecolor{hgfpale}{RGB}{236, 251, 253}
\colorlet{hgfpale10}{hgfpale!10!white}
\colorlet{hgfpale20}{hgfpale!20!white}
\colorlet{hgfpale30}{hgfpale!30!white}
\colorlet{hgfpale40}{hgfpale!40!white}
\colorlet{hgfpale50}{hgfpale!50!white}
\colorlet{hgfpale60}{hgfpale!60!white}
\colorlet{hgfpale70}{hgfpale!70!white}
\colorlet{hgfpale80}{hgfpale!80!white}
\colorlet{hgfpale90}{hgfpale!90!white}

\definecolor{hgfaerospace}{RGB}{80, 200, 170}
\definecolor{hgfast}{named}{hgfaerospace}

\colorlet{hgfaerospace10}{hgfaerospace!10!white}
\colorlet{hgfaerospace20}{hgfaerospace!20!white}
\colorlet{hgfaerospace30}{hgfaerospace!30!white}
\colorlet{hgfaerospace40}{hgfaerospace!40!white}
\colorlet{hgfaerospace50}{hgfaerospace!50!white}
\colorlet{hgfaerospace60}{hgfaerospace!60!white}
\colorlet{hgfaerospace70}{hgfaerospace!70!white}
\colorlet{hgfaerospace80}{hgfaerospace!80!white}
\colorlet{hgfaerospace90}{hgfaerospace!90!white}

\colorlet{hgfast10}{hgfast!10!white}
\colorlet{hgfast20}{hgfast!20!white}
\colorlet{hgfast30}{hgfast!30!white}
\colorlet{hgfast40}{hgfast!40!white}
\colorlet{hgfast50}{hgfast!50!white}
\colorlet{hgfast60}{hgfast!60!white}
\colorlet{hgfast70}{hgfast!70!white}
\colorlet{hgfast80}{hgfast!80!white}
\colorlet{hgfast90}{hgfast!90!white}

\definecolor{hgfearthandenvironment}{RGB}{50, 100, 105}
\definecolor{hgfee}{named}{hgfearthandenvironment}

\colorlet{hgfearthandenvironment10}{hgfearthandenvironment!10!white}
\colorlet{hgfearthandenvironment20}{hgfearthandenvironment!20!white}
\colorlet{hgfearthandenvironment30}{hgfearthandenvironment!30!white}
\colorlet{hgfearthandenvironment40}{hgfearthandenvironment!40!white}
\colorlet{hgfearthandenvironment50}{hgfearthandenvironment!50!white}
\colorlet{hgfearthandenvironment60}{hgfearthandenvironment!60!white}
\colorlet{hgfearthandenvironment70}{hgfearthandenvironment!70!white}
\colorlet{hgfearthandenvironment80}{hgfearthandenvironment!80!white}
\colorlet{hgfearthandenvironment90}{hgfearthandenvironment!90!white}

\colorlet{hgfee10}{hgfee!10!white}
\colorlet{hgfee20}{hgfee!20!white}
\colorlet{hgfee30}{hgfee!30!white}
\colorlet{hgfee40}{hgfee!40!white}
\colorlet{hgfee50}{hgfee!50!white}
\colorlet{hgfee60}{hgfee!60!white}
\colorlet{hgfee70}{hgfee!70!white}
\colorlet{hgfee80}{hgfee!80!white}
\colorlet{hgfee90}{hgfee!90!white}

\definecolor{hgfenergy}{RGB}{255, 210, 40}

\colorlet{hgfenergy10}{hgfenergy!10!white}
\colorlet{hgfenergy20}{hgfenergy!20!white}
\colorlet{hgfenergy30}{hgfenergy!30!white}
\colorlet{hgfenergy40}{hgfenergy!40!white}
\colorlet{hgfenergy50}{hgfenergy!50!white}
\colorlet{hgfenergy60}{hgfenergy!60!white}
\colorlet{hgfenergy70}{hgfenergy!70!white}
\colorlet{hgfenergy80}{hgfenergy!80!white}
\colorlet{hgfenergy90}{hgfenergy!90!white}

\definecolor{hgfhealth}{RGB}{210, 50, 100}

\colorlet{hgfhealth10}{hgfhealth!10!white}
\colorlet{hgfhealth20}{hgfhealth!20!white}
\colorlet{hgfhealth30}{hgfhealth!30!white}
\colorlet{hgfhealth40}{hgfhealth!40!white}
\colorlet{hgfhealth50}{hgfhealth!50!white}
\colorlet{hgfhealth60}{hgfhealth!60!white}
\colorlet{hgfhealth70}{hgfhealth!70!white}
\colorlet{hgfhealth80}{hgfhealth!80!white}
\colorlet{hgfhealth90}{hgfhealth!90!white}

\definecolor{hgfinformation}{RGB}{160, 35, 90}
\definecolor{hginfo}{named}{hgfinformation}

\colorlet{hgfinformation10}{hgfinformation!10!white}
\colorlet{hgfinformation20}{hgfinformation!20!white}
\colorlet{hgfinformation30}{hgfinformation!30!white}
\colorlet{hgfinformation40}{hgfinformation!40!white}
\colorlet{hgfinformation50}{hgfinformation!50!white}
\colorlet{hgfinformation60}{hgfinformation!60!white}
\colorlet{hgfinformation70}{hgfinformation!70!white}
\colorlet{hgfinformation80}{hgfinformation!80!white}
\colorlet{hgfinformation90}{hgfinformation!90!white}

\colorlet{hgfinfo10}{hgfinformation!10!white}
\colorlet{hgfinfo20}{hgfinformation!20!white}
\colorlet{hgfinfo30}{hgfinformation!30!white}
\colorlet{hgfinfo40}{hgfinformation!40!white}
\colorlet{hgfinfo50}{hgfinformation!50!white}
\colorlet{hgfinfo60}{hgfinformation!60!white}
\colorlet{hgfinfo70}{hgfinformation!70!white}
\colorlet{hgfinfo80}{hgfinformation!80!white}
\colorlet{hgfinfo90}{hgfinformation!90!white}

\definecolor{hgfmatter}{RGB}{240, 120, 30}

\colorlet{hgfmatter10}{hgfmatter!10!white}
\colorlet{hgfmatter20}{hgfmatter!20!white}
\colorlet{hgfmatter30}{hgfmatter!30!white}
\colorlet{hgfmatter40}{hgfmatter!40!white}
\colorlet{hgfmatter50}{hgfmatter!50!white}
\colorlet{hgfmatter60}{hgfmatter!60!white}
\colorlet{hgfmatter70}{hgfmatter!70!white}
\colorlet{hgfmatter80}{hgfmatter!80!white}
\colorlet{hgfmatter90}{hgfmatter!90!white}

\definecolor{codegreen}{rgb}{0,0.6,0}
\definecolor{codegray}{rgb}{0.5,0.5,0.5}
\definecolor{codepurple}{rgb}{0.58,0,0.82}
\definecolor{backcolour}{rgb}{0.95,0.95,0.92}

\lstdefinestyle{mystyle}{
    backgroundcolor=\color{hgfgray10},   
    commentstyle=\color{hgfdarkblue},
    keywordstyle=\color{hgfinformation},
    numberstyle=\tiny\color{hgfgray90},
    stringstyle=\color{hgfblue},
    basicstyle=\ttfamily\footnotesize,
    breaklines=true,
    captionpos=b,
    numbers=left,
    numbersep=5pt,
    showspaces=false,
    showstringspaces=false,
    showtabs=false,
    tabsize=2,
    aboveskip=5pt,
    belowskip=5pt,
    frame=topline,
    framerule=0pt
}
\begin{document}

\title{pyGinkgo: A Sparse Linear Algebra Operator Framework For Python}

\author{Keshvi Tuteja}
\affiliation{%
  \institution{Scientific Computing Center, Karlsruhe Institute of Technology}
  \country{Germany}
}
\email{keshvi.tuteja@kit.edu}

\author{Gregor Olenik}
\affiliation{%
  \institution{Computational Mathematics, Technical University of Munich}
  \country{Germany}
}
\email{gregor.olenik@tum.de}

\author{Roman Mishchuk}
\affiliation{%
  \institution{Technical University of Munich}
  \country{Germany}
}
\email{roman.mishchuk@tum.de}

\author{Yu-Hsiang Tsai}
\affiliation{%
  \institution{Computational Mathematics, Technical University of Munich}
  \country{Germany}
}
\email{yu-hsiang.tsai@tum.de}

\author{Markus G{\"o}tz}
\affiliation{%
  \institution{Helmholtz AI and SCC, Karlsruhe Institute of Technology}
  \country{Germany}
}
\email{markus.goetz@kit.edu}

\author{Achim Streit}
\affiliation{%
  \institution{Scientific Computing Center, Karlsruhe Institute of Technology}
  \country{Germany}
}
\email{achim.streit@kit.edu}

\author{Hartwig Anzt}
\affiliation{%
  \institution{Technical University of Munich and University of Tennessee}
  \country{Germany, USA}
}
\email{hartwig.anzt@tum.de}

\author{Charlotte Debus}
\affiliation{%
  \institution{Scientific Computing Center, Karlsruhe Institute of Technology}
  \country{Germany}
}
\email{charlotte.debus@kit.edu}

\renewcommand{\shortauthors}{Tuteja et al.}

\begin{abstract}
Sparse linear algebra is a cornerstone of many scientific computing and machine learning applications. Python has become a popular choice for these applications due to its simplicity and ease of use. Yet high-performance sparse kernels in Python remain limited in functionality, especially on modern CPU and GPU architectures. We present pyGinkgo, a lightweight and Pythonic interface to the Ginkgo library, offering high-performance sparse linear algebra support with platform portability across CUDA, HIP, and OpenMP backends. pyGinkgo bridges the gap between high-performance C++ backends and Python usability by exposing Ginkgo’s capabilities via Pybind11 and a NumPy and PyTorch compatible interface. We benchmark pyGinkgo’s performance against state-of-the-art Python libraries including SciPy, CuPy, PyTorch and TensorFlow. Results across hardware from different vendors demonstrate that pyGinkgo consistently outperforms existing Python tools in both Sparse Matrix Vector (SpMV) product and iterative solver performance, while maintaining performance parity with native Ginkgo C++ code. Our work positions pyGinkgo as a compelling backend for sparse machine learning models and scientific workflows.
\end{abstract}

\maketitle

\begingroup
\renewcommand\thefootnote{}\footnotetext{%
\textit{Preprint.} Accepted for publication at the 54\textsuperscript{th} International Conference on Parallel Processing (ICPP’25). 
The final published version is available at 
\href{https://doi.org/10.1145/3754598.3754648}{https://doi.org/10.1145/3754598.3754648}.}
\addtocounter{footnote}{0}
\endgroup

\section{Introduction}

Sparse linear algebra plays a crucial role in many scientific and engineering applications, including fluid dynamics, heat transfer problems, and data mining. Frameworks and libraries used for high-performance sparse linear algebra are typically very advanced and offer a highly optimized stack of computational kernels~\cite{Cusp, magma}. However, these are mostly implemented in low-level languages like C++ and Fortran. 

While using low-level languages has obvious performance benefits, Python has emerged as a popular choice for scientific computing and data science applications due to its ease of use, readability, and extensive ecosystem of libraries. Its dominance is especially visible in the machine learning community. In contrast to its strengths in dense linear algebra and general usability, Python’s ecosystem for scalable and efficient sparse computations remains relatively underdeveloped.
For example, SciPy~\cite{scikit-learn} and NumPy~\cite{harris2020array}, rely on frameworks like BLAS~\cite{10.1145/77626.77627}~\cite{10.1145/567806.567807}~\cite{10.1145/77626.79170} and LAPACK~\cite{laug} for dense linear algebra operations, but the functionality and hardware support for sparse linear algebra remains limited. 

The need for sparse linear algebra libraries in Python is currently becoming even more imminent, given the intensive developments towards sparse neural networks~\cite{hoefler2021sparsity}. These sparse models, e.g., when most elements of the weight matrices tend to zero, rely heavily on core operations like sparse matrix-vector and matrix-matrix products.  Even though the prevailing workloads, such as Transformers~\cite{vaswani2023attentionneed}, are using dense representations, there are a growing number of methods that leverage sparse formulation, including, but not limited to, spiking and graph neural networks as well as the weight matrices after the gradient-descent based optimization~\cite{collins2014memoryboundeddeepconvolutional, 10.3389/fnins.2022.760298, srinivas2016trainingsparseneuralnetworks}. 

Ginkgo~\cite{ginkgo-toms-2022} is a C++ based high-performance linear algebra library with a special focus on sparse linear systems. To bridge the performance gap between dense and sparse computations in the Python world, we present pyGinkgo -- Python bindings for the Ginkgo library. pyGinkgo utilizes Pybind11~\cite{jakob2017pybind11} to access Ginkgo's C++ kernels from Python side, which provides a Pythonic API as well as interoperability with NumPy and PyTorch, while introducing only a minimal performance overhead. This design enables Python users to leverage Ginkgo’s advanced capabilities for performing sparse computations, offering significant potential for improving the performance in python-based scientific computing and machine learning workloads. 

Our key contributions through this work are as follows:
\begin{itemize}
    \item pyGinkgo, a sparse linear algebra library for Python, providing access to Ginkgo's powerful iterative solvers and SpMV kernels.
    \item A thorough performance evaluation of state-of-the-art sparse linear algebra frameworks in Python, focusing on matrix-vector multiplication (SpMV) and iterative solvers. The comparison spans both CPU and GPU platforms, including hardware from NVIDIA and AMD.
    \item An empirical estimation of the engineering overhead introduced by Pybind11~\cite{jakob2017pybind11} when comparing pyGinkgo to the native C++ implementation of Ginkgo, demonstrating marginal performance loss.
\end{itemize}
The implementation of pyGinkgo is publicly available at \textcolor{blue}{\url{https://github.com/Helmholtz-AI-Energy/pyGinkgo}}.

\section{Related Work and State-of-the-Art}
\label{sec:Related Work}

Over the years, several efforts have been made to enable sparse linear algebra capabilities in Python. Early developments in this area mainly relied on low-level backends originally designed for C and C++ environments. Some notable examples implemented in low-level languages are Ginkgo, libsrb~\cite{libsrb}, cuSPARSE~\cite{cusparse}, MAGMA~\cite{magmasparse} and Kokkos Kernels~\cite{kokkos}. Among these resources, we selected Ginkgo due to its combination of modern design, performance portability across multiple hardware, and rich support for iterative solvers and preconditioners.

Building on these low-level foundations, a number of high-level Python libraries have emerged to improve usability and integration within the Python ecosystem. These libraries either wrap existing low-level backend or offer native sparse functionality. Some notable examples are SLEPc~\cite{slepc}, PyTrilinos~\cite{PyTrilinos}, FEniCS~\cite{AlnaesEtal2014}, PyAMG~\cite{pyamg2023} and PySparse~\cite{pysparse-docs}. These libraries provide varying degrees of sparse matrix support, often targeting specialized applications such as finite element analysis, eigenvalue problems and algebraic multigrid methods. However, most of these libraries either lack GPU acceleration or compatibility with modern hardware, fast implementations, or a Pythonic interface. As a consequence, some of these libraries have become obsolete or are no longer actively maintained.

More recently, major Python frameworks such as PyTorch~\cite{paszke2019pytorch}, TensorFlow~\cite{tensorflow2015-whitepaper}, SciPy, and CuPy~\cite{cupy_learningsys2017} have progressively incorporated support for sparse linear algebra operations. PyTorch and TensorFlow provide support for sparse matrices and a limited set of sparse operations and its integration with neural networks is restricted. Both of these libraries lack advanced features like iterative solvers and preconditioners. TensorFlow only supports the Coordinate (COO) matrix format. Moreover, computations at double precision in PyTorch and TensorFlow are rather inefficient, making them unsuitable for many high-performance computing tasks. Our own results show that their SpMV kernel is not optimized (see section \Cref{Res:res}).
In comparison, CuPy and SciPy offer decent support. They provide functionality for various sparse matrix formats as well as various direct and iterative solvers. CuPy leverages several CUDA Toolkit libraries such as cuBLAS~\cite{cublas}, cuRAND~\cite{crand}, cuSOLVER~\cite{cusolver}, cuSPARSE, cuFFT~\cite{cufft}, cuDNN~\cite{chetlur2014cudnnefficientprimitivesdeep}, and NCCL~\cite{cunccl} to provide functionality similar to SciPy. Although these contemporary libraries offer broad dense computation capabilities, most lack comprehensive support for sparse matrices, iterative solvers, or efficient parallelism on both CPUs and GPUs. This highlights the need for an optimized, portable and extensible sparse linear algebra library like pyGinkgo.


\section{Overview}
\label{Theory:overview}


\begin{figure}
    \centering
    \input{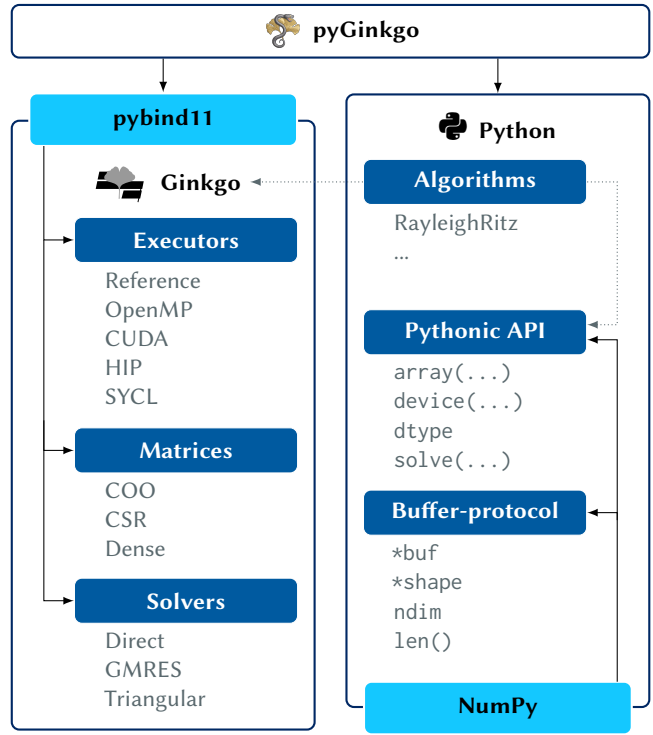}
    \caption{Overview of pyGinkgo}
    \label{fig:pyGinkgo}
\end{figure}

pyGinkgo is built as a wrapper around Ginkgo. Its design is rooted in three main building blocks:
\begin{itemize}
    \item The C++ Core: Ginkgo’s high-performance algorithms and data structures are implemented in C++ and it is the central building block for pyGinkgo.
    \item Python Bindings: Pybind11 is used to create a thin wrapper around Ginkgo enabling the support for all the accelerator backends already provided by Ginkgo
    \item A user-friendly Python Application Programming Interface (API) that provides basic interoperability with commonly used Python packages. 
\end{itemize}

\subsection{Principal Architecture of pyGinkgo}
The pyGinkgo library consists of two principal parts:
\begin{enumerate}
\item The C++ bindings using pybind11~\cite{jakob2017pybind11} that generate functions and classes which are accessible from Python. 
\item Further implementations in pure Python, which are responsible for a seamless integration into the existing Python ecosystem by providing complex dispatching based on data-types or implementing complex mathematical algorithms. This includes integration with NumPy and SciPy and PyTorch interoperability to make pyGinkgo more idiomatic to Python.
\end{enumerate}
\Cref{fig:pyGinkgo} shows the general structure of the pyGinkgo library and its main features. In addition to exposing functionality provided by Ginkgo, pyGinkgo also includes several algorithms and methods implemented purely in Python. Currently, we provide implementations of the Rayleigh–Ritz method, with ongoing development to expand this further. We also provide methods like \verb|array()| and \verb|device()| to make the API more intuitive. By design, pyGinkgo tries to implement a Pythonic interface. For example, operations on matrices and vectors resemble NumPy idioms, and solver pipelines can be constructed with minimal boilerplate code. This balance between leveraging Ginkgo’s performance and Python’s simplicity is central to pyGinkgo’s design philosophy. Note, that pyGinkgo is still under active development. We plan to add support for various other matrix formats and algorithms that Ginkgo offers.  

\subsection{Ginkgo as a Computational Engine}\label{sec:ginkgo_backend}
Ginkgo is a high-performance linear algebra library with a focus on sparse linear systems. It provides high-performance algorithms and data structures implemented in C++~\cite{ginkgo-toms-2022, GinkgoJoss2020}.
One of its primary design choices is platform-portable performance~\cite{tsai2020amdspmv, anzt2020spmv, 9307857}, ensuring that computations run optimally across a diverse range of hardware architectures from major vendors. To accomplish this, Ginkgo implements its computational kernels using vendor-native programming models, such as CUDA for NVIDIA GPUs, HIP for AMD GPUs, and SYCL for Intel GPUs, while also leveraging OpenMP annotated C++ code for efficient multi-threaded execution on CPUs. This approach enables Ginkgo to take full advantage of hardware-specific optimizations while maintaining a unified and consistent interface for users. By providing a common C++ API, Ginkgo abstracts away the complexity of hardware-specific implementations, allowing developers to focus on their applications without having to modify their code for different computing platforms. While Ginkgo is implemented in C++ for high performance, many users prefer Python due to its ease of use, extensive ecosystem, and strong support for numerical computing through libraries like NumPy and SciPy. For example, libraries such as TensorFlow, PyTorch, and JAX~\cite{jax2018github} also offer integration with NumPy and SciPy. These libraries have gained widespread popularity because they successfully combine a Pythonic interface with high-performance C/C++ backends. These libraries follow a common design philosophy: expose minimal and idiomatic APIs to the user while deferring performance-critical operations to lower-level code. pyGinkgo is built on the same underlying design principles.
However, while these libraries have focused primarily on dense operations, pyGinkgo addresses this gap. This would be especially beneficial for the applications that require efficient sparse matrix computations and GPU-accelerated operations, as Ginkgo's optimized kernels for CUDA, HIP, and SYCL could be directly utilized from Python.

\subsection{Pybind11 for Python Bindings}
Since Ginkgo implements a C++ API to access its algorithms and data structures, a way to interact with the C++ API from Python code is required. When selecting a tool to create Python bindings for the Ginkgo C++ codebase, we had the following key requirements:
\begin{itemize}
    \item Minimal performance overhead
    \item Pythonic API and interoperability with NumPy
    \item Modern C++ support
    \item Simple build integration
\end{itemize}
After evaluating several options, we found that pybind11 was the best fit for our needs, especially when compared to alternatives like SWIG~\cite{beazley1996swig}, Boost.Python~\cite{abrahams2003building}, Cython~\cite{behnel2011cython}, and ctypes. Pybind11's modern, lightweight, and header-only design enables the generation of high-performance Pythonic APIs with minimal boilerplate, avoiding the complexity of intermediate layers. Unlike SWIG, which supports multiple languages at the cost of increased overhead and less intuitive bindings, pybind11 focuses exclusively on seamless C++/Python interoperability, providing direct and natural mappings for C++ classes, templates, and STL containers.
We also considered Cython, which is known for its ease of use and rapid development, especially for calling simple C functions or writing performance-critical Python code. However, Cython falls short when dealing with more advanced C++ features, such as class hierarchies, templates, and lifetime management of native objects. Moreover, Pybind11 includes valuable features like NumPy support, keyword arguments, and smart pointer management.
Although Boost.Python offers similarly powerful functionality, but it comes at the price of heavier dependencies, slower compilation, and more complex setups. Since we are focusing on performance-critical workloads, pybind11 was the superior choice. As demonstrated in section \Cref{section:pgko_vs_gko}, pyGinkgo is quite light-weight in terms of relative performance difference with respect to Ginkgo's native implementation.

\subsection{Algorithms in Pure Python}
Ginkgo is built entirely in C++ and relies heavily on C++ features to enable modular and efficient interactions between its components. However, experimenting with new ideas or customizing algorithms within a C++ codebase can pose a significant barrier for users who are unfamiliar with advanced C++ or are more comfortable with Python. With pyGinkgo, users can now explore new methods and integrate Ginkgo into their applications directly from Python, making it much easier to adopt high-performance linear algebra in diverse scientific workflows.

To ensure pyGinkgo feels like a native Python library rather than a thin C++ wrapper, additional care is needed beyond the use of pybind11. For example, Python has no direct equivalent to function overloading, which we discuss more in \Cref{sec:Types}, and instead accepts types indiscriminately as arguments to a function if no explicit type checking is performed. Thus, naturally, Python APIs are different compared to C++ APIs, often with a single entry point function that implements complex dispatching mechanisms based on the arguments passed. For an integration into the Python ecosystem, pyGinkgo implements comparable entry point functions, such as \verb|as_tensor|, \verb|solve|, or \verb|device| in PyTorch or CuPy, that require a dispatching mechanism to the functions provided by pybind11. As proof of concept, we implemented Rayleigh-Ritz method on the Python side, which is not natively supported by Ginkgo yet. This demonstrates the extensibility of our framework, through which users can construct complex algorithms composed of existing linear algebra operations exposed using pyGinkgo. For instance, Rayleigh–Ritz relies on repeated applications of operations such as sparse matrix–vector products and other operations which are available as Ginkgo operators. This allows users to focus on algorithm design without worrying about low-level GPU or CPU parallelization details. Hence, prototyping of new solvers or algorithmic variants is possible, while still leveraging the performance and portability of the Ginkgo backend.
\vspace{-1em}
\subsection{pyGinkgo API} 
In this section, we discuss how to use the pyGinkgo API to solve a sparse linear system of the form $Ax=b$ as shown in \Cref{lst:usage_specific}. In principle, pyGinkgo provides two ways to create a linear solver: first, via specific solver bindings shown in \Cref{lst:usage_specific} and second, via a generic solver interface called the \texttt{config solver}.

Initially, a CUDA device (\verb|pg.device('CUDA')|) is instantiated to enable GPU-based computations. The sparse matrix in Compressed Sparse Row (CSR) format is then loaded using the \verb|read| function. The input matrix is read from a Matrix Market (MTX) file named\ \verb|m1.mtx|. This is demonstrated in lines \hyperref[line:usage_config_mtx_cuda]{4-7}. The \verb|device| function is an abstraction for the \verb|Executor| on Ginkgo's side. It determines the device on which data is stored and the computations are performed. As consistent with Ginkgo, pyGinkgo currently supports CUDA, HIP, and OpenMP executors (discussed in \Cref{sec:exec}). 

The right-hand side vector b and the solution vector x are initialized with \verb|double| (\verb|np.float64|) for numerical precision. pyGinkgo currently supports half, single and double-precision values and index types (discussed in \Cref{sec:Types}). 

Vectors can be instantiated directly using \verb|pyGinkgo.as_tensor| by specifying the executor, shape, and fill value, as demonstrated for \verb|b| and \verb|x| in lines \hyperref[line:usage_config_righthand]{9-14}. Alternatively, a NumPy array can be created and copied into a pyGinkgo tensor (discussed in \Cref{sec:buffer_protocol}).

After instantiating the necessary data structures, \verb|mtx|, \verb|x|, and \verb|b|, an \verb|Ilu| preconditioner is instantiated. In the example, generalized minimal residual method (GMRES), a Krylov subspace method is instantiated with a Krylov dimension of 30. The solver is configured to stop based on a maximum of 1000 iterations or a relative residual reduction factor of $10^{-6}$. This is specified in lines \hyperref[line:usage_config_precond]{16-22}. The \verb|apply| method is then called. The function returns a logger, which provides diagnostic information about convergence and iteration progress, and the solution vector, which overwrites the initial guess.

In addition to providing the direct bindings to Ginkgo's solver classes, pyGinkgo also offers a configuration-based interface. When the \verb|solve| method is called on pyGinkgo's side, a dictionary that is based on the arguments that are passed is created at the python backend. An example dictionary is as seen in lines \hyperref[line:usage_config_args]{1-16} in \Cref{lst:usage_config}. This dictionary is then used to call Ginkgo's  \verb|config_solve| method. In the example, GMRES is instantiated with a Krylov dimension of 30 and a Jacobi preconditioner with a block size of 1. The solver is configured to stop based on a maximum of 1000 iterations or a relative residual reduction factor of $10^{-6}$. 

\begin{flushright}
\begin{minipage}{\linewidth}
\begin{lstlisting}[language=Python, style=myStyle, label=lst:usage_specific, caption= \normalfont Python example demonstrating the use of pyGinkgo to solve a sparse linear system using the GMRES iterative solver with an ILU preconditioner using the direct solver bindings.,firstnumber=1,escapechar=|]
import pyGinkgo as pg
import numpy as np

fn = 'm1.mtx'|\label{line:usage_config_mtx_cuda}|
dev = pg.device("cuda") 
mtx = pg.read(device=dev, path=fn, dtype="double", format="Csr")
n_rows = mtx.size[0]

b = pg.as_tensor(|\label{line:usage_config_righthand}|
  device=dev, dim=(n_rows,1), dtype="double", fill=1.0
)
x = pg.as_tensor(
  device=dev, dim=(n_rows,1), dtype="double", fill=0.0
)

# Create ILU preconditioner|\label{line:usage_config_precond}|
preconditioner = pg.preconditioner.Ilu(dev, mtx)

#Setup GMRES solver
solver = pg.solver.gmres(dev, mtx, preconditioner,
    max_iters=1000, krylov_dim=30, reduction_factor=1e-06
)

#Apply
logger, result = solver.apply(b, x)

\end{lstlisting}
\end{minipage}
\end{flushright}

\begin{flushright}
\begin{minipage}{\linewidth}
\begin{lstlisting}[language=Python, style=myStyle, label=lst:usage_config, caption=\normalfont An example dictionary created at Python backend to be passed to Ginkgo's solver kernels demonstrating the use of pyGinkgo to solve a sparse linear system using the GMRES iterative solver with a Jacobi preconditioner on a CUDA executor using the config solver.,escapechar=|]
args = {|\label{line:usage_config_args}|
    "type": "solver::Gmres",
    "krylov_dim": 30,
    "preconditioner": {
        "type": "preconditioner::Jacobi",
        "max_block_size": 1
    },
    "criteria": [
        {"type": "Iteration", "max_iters": 1000},
        {
          "type": "ResidualNorm", 
          "reduction_factor": 1e-6, 
          "baseline": "rhs_norm"
        }
    ],
}

\end{lstlisting}
\end{minipage}
\end{flushright}

\section{Implementation Details and Advanced Concepts}\label{Theory:Implementation}

\subsection{Device Access via Executor \label{sec:exec}}
pyGinkgo provides bindings for Ginkgo's executor class that determines where the underlying data is stored and where computations take place. Executors play a crucial role in managing memory and controlling execution across different hardware backends. These are also responsible for allocation/de-allocation of memory on device, synchronizing computations between host and device as well as copying data between multiple executors. pyGinkgo relies on pybind11's support for smart pointers allowing Python to share ownership with C++ in a safe way.
As already discussed in section \ref{sec:ginkgo_backend}, pyGinkgo currently implements bindings to the CUDA, HIP, Omp, and reference executors offered by Ginkgo.
For integration of pyGinkgo into the existing ecosystem, pyGinkgo refers to the executor as \verb|device|. Additionally, pyGinkgo provides a \verb|pyGinkgo.device(name, id=0)| factory function to create an instance of an executor. Functions typically accept an executor instance as an argument, and data structures provide a getter to retrieve an assigned executor.
Executor objects are among the first components to be defined when writing code with pyGinkgo. It is important to note that a program can utilize multiple executors simultaneously, allowing users to perform different operations and store data on specific devices as needed. Using multiple executors may have additional overhead in copying the data from one device to another. 

To better illustrate some of the pybind11 concepts, the bindings for the executor base class and the CUDA executor are shown in the \Cref{lst:exec}. The binding to a C++ class is declared by \verb|pybind|'s \verb|class_<T>(module, "name")|. In the given syntax, \verb|T| are a set of template arguments, where typical arguments are the corresponding class for which bindings are generated, any base classes of the class, and the holder type. For more information on further template parameters, we refer to the pybind11 documentation\footnote{\url{https://pybind11.readthedocs.io/en/stable/advanced/classes.html}}. Specifying information about the base class allows to then pass any child classes as arguments, which accepts the base class as a parameter to generated functions. The holder type defines whether the underlying instance can be accessed as a reference or via a smart pointer, an important aspect regarding the fact that Ginkgo handles objects with smart pointers. This behavior is enforced by implementing constructors as protected, thus forcing the instantiation of concrete objects via static create functions, which, in turn, always return a smart pointer to the created object. 

Furthermore, the function argument \verb|module| defines the Python module in which the corresponding Python class should be registered, and \verb|name| defines the name of the class on the Python side. In this example, it is \verb|CUDA|.
\begin{lstlisting}[language=C++, style=myStyle, label=lst:exec, caption= \normalfont Exemplary implementation of the CudaExecutor binding.]
py::class_<gko::CudaExecutor,
    gko::detail::ExecutorBase<gko::CudaExecutor>,
    std::shared_ptr<gko::CudaExecutor>
>(module, /*name = */"CUDA")
    .def(py::init([](int dev_id,
           std::shared_ptr<gko::Executor> master,
           std::shared_ptr<gko::CudaAllocatorBase> alloc,
           std::shared_ptr<gko::cuda_stream> stream
        ) {
            return gko::CudaExecutor::create(dev_id, master, alloc, stream->get());
        }));
\end{lstlisting}
\subsection{Ginkgo's Linear Operator Abstraction\label{sec:LinOp}}
The \texttt{LinOp} (Linear Operator) class is a core abstraction in Ginkgo, serving as the base type for nearly all data structures, including matrices, solvers, and preconditioners. Each concrete implementation of \texttt{LinOp} defines an \texttt{apply} method, which specifies how the operator acts on input vectors. Under this abstraction lies a variety of operations: a matrix \texttt{LinOp} performs a SpMV, a solver \texttt{LinOp} applies a linear system solver to a right-hand side vector (as seen in \Cref{lst:usage_specific}), and a preconditioner \texttt{LinOp} applies a transformation to accelerate convergence. With this abstraction, every object that models a linear operation can be applied using the same method call. This unified interface fosters a high degree of composability, enabling users to construct sophisticated solver pipelines by combining different linear operators. \texttt{pyGinkgo} preserves this design philosophy, extending the \texttt{LinOp} abstraction and its capabilities to Python users.
\section{Generic Ginkgo Solver Entry Points}
\begin{figure}[b]
    \centering
    \resizebox{0.95\linewidth}{!}{\input{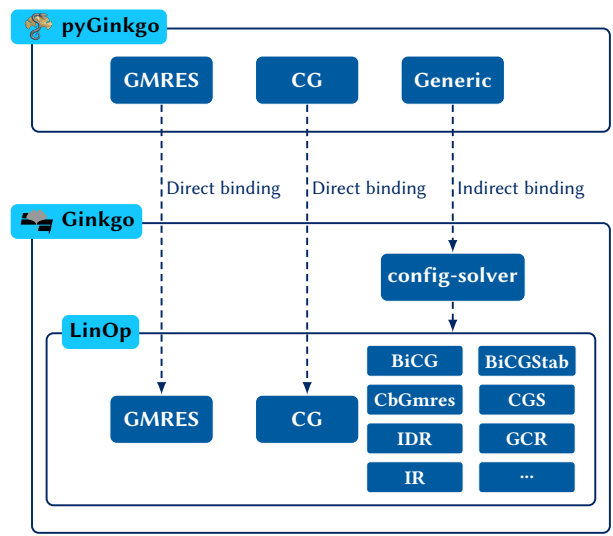}}
    \caption{Overview of Solver Bindings}
    \label{fig:overview_solver}
\end{figure}

Ginkgo provides a generic entry point for its solver via configuration parameters in JSON, YAML, or custom format. This generic solver entry point, internally also referred to as config-solver, can be instantiated as any other LinOp via a LinOpFactory and invoked through the LinOp's apply function, as discussed in \Cref{sec:LinOp}. With this method, users of Ginkgo gain access to all available solvers and specialized preconditioners without having to write code for each use case. Additionally, this also allows the selection of solvers and preconditioner combinations at run-time, thus giving users the flexibility to experiment with different setups by simply modifying a configuration without having to recompile the application. 

pyGinkgo leverages this approach to gain access to a rich set of solvers, factorizations, and preconditioners without the need to explicitly bind them. Additionally, this reduces the development pressure on the pyGinkgo side: whenever Ginkgo introduces a new functionality, which can be accessed via the generic solver entry point, pyGinkgo users only need to update to a recent version of Ginkgo without requiring explicit bindings. While the typical use case is to read the configuration from a file, pyGinkgo also implements a wrapper for the generic entry point that generates the configuration parameters in JSON format from Python's dictionary directly as also shown earlier in \Cref{lst:usage_config}, without depending on any temporary configuration files on disk. 

A drawback of the configuration file approach is that the configuration files themselves can become difficult to implement and to verify, since currently no JSON schema for validation is available. Additionally, for developers, interacting with the data structures directly instead of writing JSON configurations can be preferable because of features like auto-completion or access to doc-strings of the classes and functions through Python's \verb|help()| and \verb|dir()| functions. Thus, pyGinkgo provides access to Ginkgo's solvers via the config-solver interface. Additionally, it also provides explicit bindings to commonly used solvers like GMRES, the direct solver, and triangular solvers, and to preconditioners like IC and ILU, which is illustrated in \Cref{fig:overview_solver}. We plan to expand this set further.

\subsection{Available Value and Index Types\label{sec:Types}}
To implement datatypes and algorithms in Ginkgo that support lower, higher, and mixed precision, Ginkgo relies heavily on C++ templates.
However, the concept of templated types and the corresponding potential overloading of functions is not directly applicable to Python. For example, the functions \verb|funcxx(int a)| and \verb|funcxx(float a)| can have completely different implementations, whereas in Python, only a single \verb|funcxx(a)| can exist. This issue stems from the language differences, and is resolved in practice via pre-instantiation of all possible template parameter combinations that the Python side might require. In pyGinkgo, when creating bindings to overloaded functions, the functions are disambiguated by appending the typename to the function, resulting in \verb|funcxx_int(a)| and \verb|funcxx_float(a)|. 

This raises the issue of abstracting away from these parameters on the bindings side and simplifying the access for the users, as is done e.g. in NumPy or PyTorch. They implement both raw and abstracted away functions on C/C++ side. Our current approach consists of directly exposing instantiated C++ templates inside of \verb|pyGinkgo.pyGinkgoBindings| module and then using them within the more abstracted Python API inside of \texttt{pyGinkgo} module. Thus, a function named \verb|funcxx(a)| is implemented in the pyGinkgo module, which simply dispatches to either
\verb|funcxx_int(a)| or \verb|funcxx_float(a)| based on the type of a.

In comparison to NumPy or PyTorch source code, Ginkgo has more templated parameters, which results in a very large number of instantiations. Thus, the decision to introduce a second layer of abstraction on the Python side originates from the fact that implementing it in C++ would have provided little to no performance benefits while drastically increasing the complexity and amount of code. Additionally, it allows for easier implementation of more sophisticated user algorithms on the Python side by utilizing direct access to the exposed C++ API.

Currently, pyGinkgo supports three value and two index types, which are listed in \Cref{tab:dtypes}.

\begin{table}
\caption{Overview of the available data and index types}\label{tab:dtypes}
\centering
\begin{tabular}{
     @{\hspace*{3px}}
    c @{\hspace*{3px}}  @{\hspace*{3px}}
    l @{\hspace*{3px}}  @{\hspace*{3px}}
    c
    @{\hspace*{3px}} 
}
    \toprule
    \thead{Size (bytes)} & \thead{Value Type} & \thead{Index Type} \\
    \midrule
    2 & \texttt{half} &  \\
    4 & \texttt{float} & \texttt{int32} \\
    8 & \texttt{double} & \texttt{int64} \\
    \bottomrule
\end{tabular}
\end{table}

\subsection{Buffer-protocol and Interoperability with the Python Ecosystem}\label{sec:buffer_protocol}
Python's buffer protocol \footnote{\url{https://docs.python.org/3/c-api/buffer.html}} defines a standard for the direct access to an object's memory, without requiring a copy. This is particularly useful when integrating Python with low-level, high-performance code written in languages like C or C++. By exposing a direct memory buffer, the protocol enables efficient data sharing between Python and external libraries, improving both performance and interoperability.

NumPy supports the buffer protocol, allowing its arrays to be passed to pyGinkgo without duplication. This significantly enhances efficiency, as Python can act as a lightweight wrapper over low-level implementations. 

As a result, NumPy objects can be passed to Ginkgo on the C++ side in the following ways:
\begin{lstlisting}[language=Python] 
x = np.random.rand(dim).astype(np.float32)
x1 = pg.matrix.dense_float(executor, x) # via binding
x2 = pg.as_tensor(x, device=executor) # via dispatch
\end{lstlisting}
Interoperability is a key requirement for modern scientific software, particularly in Python, where users often construct complex workflows by combining multiple specialized libraries. A flexible interface that integrates naturally with tools like NumPy, SciPy, or PyTorch enables users to experiment and deploy algorithms without being locked into a rigid ecosystem. Like other major libraries, pyGinkgo offers very low computational overhead which makes interoperability feasible and efficient.






\begin{figure*}[!h]
    \centering
    \begin{subfigure}[b]{0.30\textwidth}
        \centering
        \includegraphics[width=\textwidth]{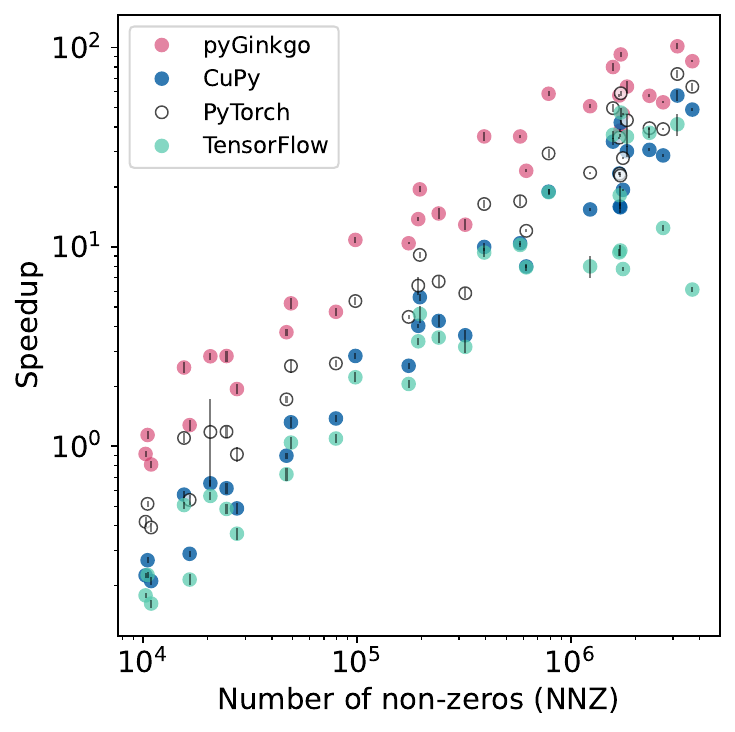}
        \caption{\small SpMV on NVIDIA A100 GPU: Variation of speedup relative to SciPy with number of non-zeros}
        \label{fig:nnz_gpu}
    \end{subfigure}
    \hfill
    \begin{subfigure}[b]{0.30\textwidth}
        \centering
        \includegraphics[width=\textwidth]{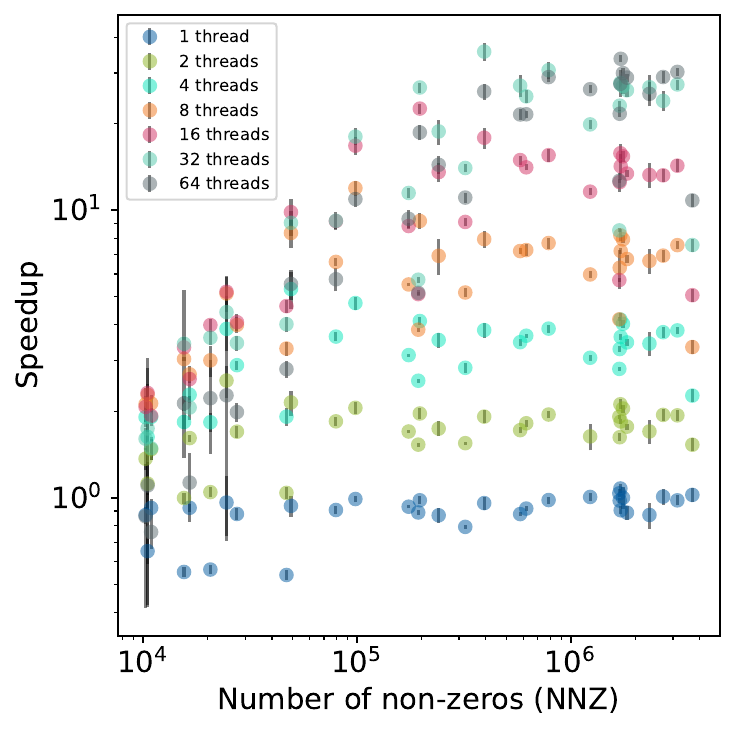}
        \caption{\small SpMV on Intel Xeon Platinum 8368 CPU: Variation of speedup relative to SciPy with number of non-zeros}
        \label{fig:nnz_cpu}
    \end{subfigure}
    \hfill
    \begin{subfigure}[b]{0.30\textwidth}
        \centering
        \includegraphics[width=\textwidth]{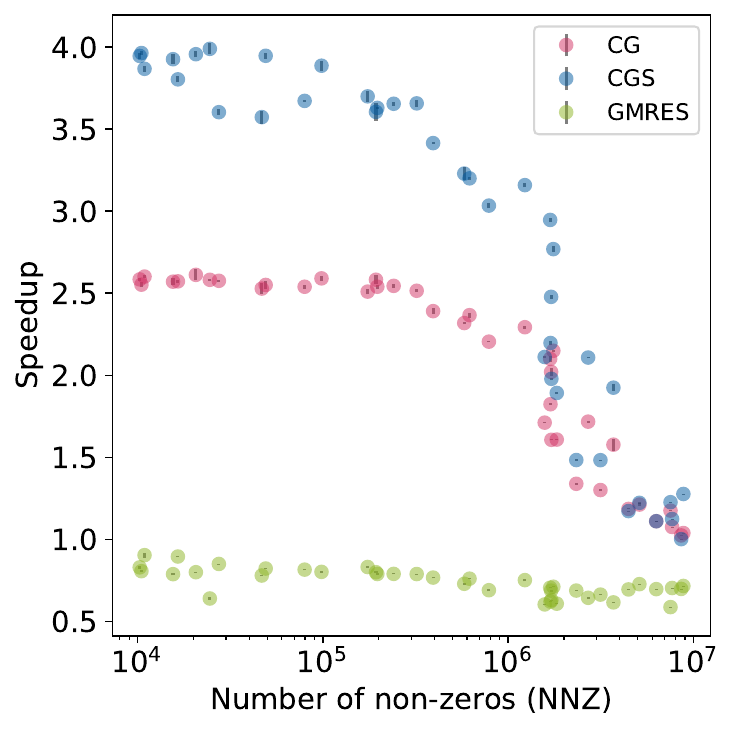}
        \caption{\small Solver on NVIDIA A100: Variation of speedup relative to CuPy with number of non-zeros for 1000 solver iterations}
        \label{fig:solver_nvidia}
    \end{subfigure}
    \caption{SpMV and solver performance on different architectures}
    \label{fig:ba_gpu}
\end{figure*}

\begin{figure*}[!h]
    \centering
    \begin{subfigure}{0.35\textwidth}
        \includegraphics[width=\textwidth]{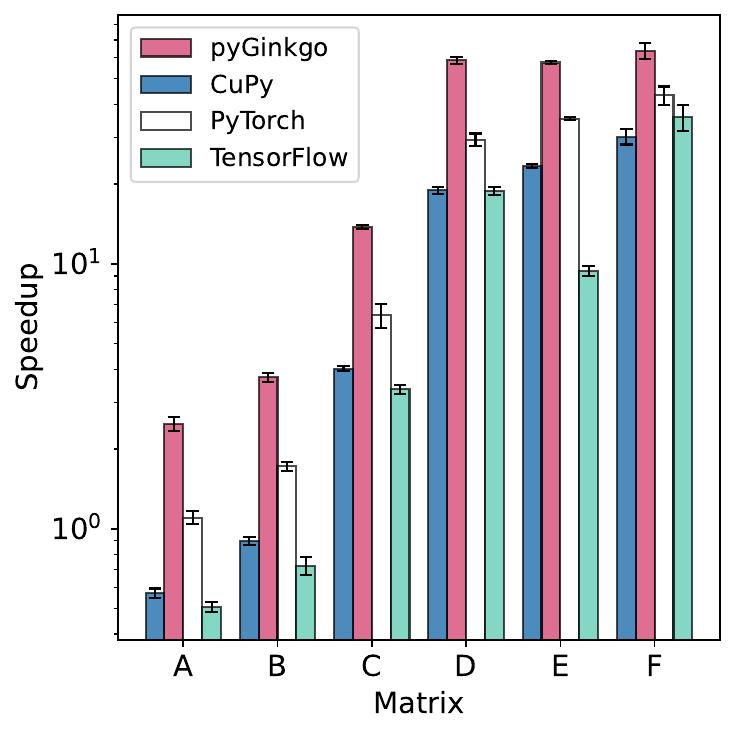}
        \caption{SpMV on NVIDIA A100 GPU}
        \label{fig:bar_gpu}
    \end{subfigure}
    \begin{subfigure}{0.35\textwidth}
        \includegraphics[width=\textwidth]{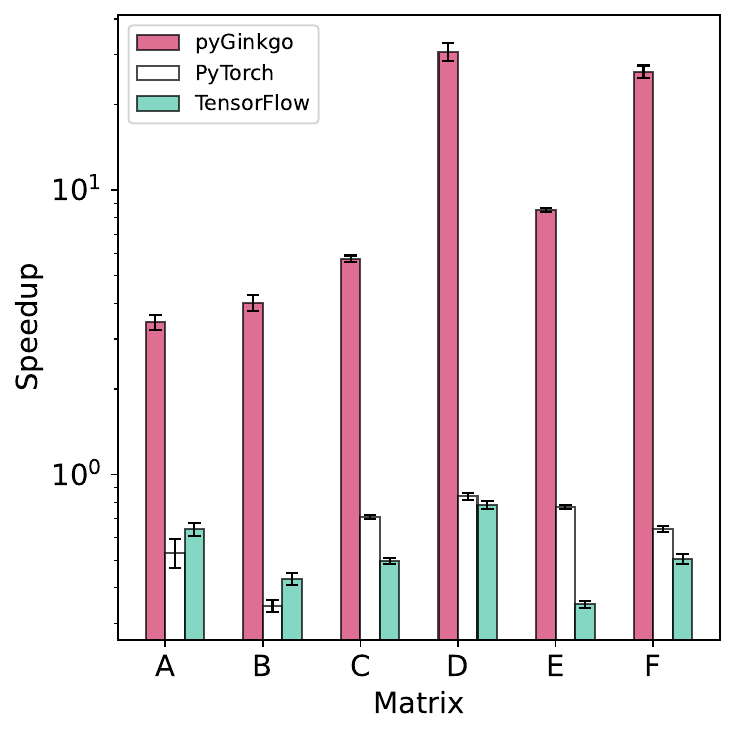}
        \caption{SpMV on Intel Xeon Platinum 8368 CPU}
        \label{fig:bar_cpu}
    \end{subfigure}
    \caption{Variation of Speedup relative to SciPy for representative matrices on GPU and CPU}
    \label{fig:bar}
\end{figure*}

\section{Performance Results and Analysis}\label{Res:res}

To demonstrate the practical benefits of pyGinkgo, we benchmarked the performance of SpMV and solver kernels across widely used Python libraries, including PyTorch, SciPy, TensorFlow, and CuPy, on both CPU and GPU platforms. Benchmarks for SpMV kernels were conducted using a dataset of 30 sparse matrices from the SuiteSparse Matrix Collection~\cite{Kolodziej2019}. Benchmarks for solver kernels were conducted using 40 sparse matrices. These test matrices cover a wide range of dimensions and densities, with dimensions up to $10^6$ and densities below $1\%$ in all cases except for five with a density greater than $1\%$. Since machine learning workloads primarily rely on SpMV in low precision, the SpMV benchmarks were conducted using single-precision. In contrast, most scientific computing workflows that require solvers demand double precision for accuracy, so the solver benchmarks were performed using double-precision. Benchmarks were conducted for both CSR and COO matrix formats. 

We present the results for the best-performing matrix formats for each library. The benchmarks were executed on the HoreKa supercomputer, featuring Intel Xeon Platinum 8368 CPUs and NVIDIA A100 GPUs. A single CPU node on HoreKa consists of 2 sockets with 76 physical cores each. We assessed single-core and multi-core performance for CPUs, alongside GPU execution performance. We also compared pyGinkgo's performance with that of Ginkgo's on a set of 45 sparse matrices from the SuiteSparse Matrix Collection. This was done on NVIDIA A100 and AMD Instinct MI100 accelerators for both CSR and COO matrix formats.

All experiments presented in this paper were conducted using Python 3.11. For the NVIDIA backend, we used CUDA 12.2, loaded via the environment module \texttt{devel/cuda/12.2}. For the AMD backend, we used ROCm 6.2.2 loaded via the environment module \texttt{toolkit/rocm/6.2.2}. These module configurations were applied consistently during benchmarking to ensure a fair comparison across hardware platforms

\subsection{SpMV}
To evaluate pyGinkgo’s performance for SpMV, we considered operations of the form $x=Ab$, where $A$ is a sparse matrix from the dataset and $b$ is initialized as a vector of random values. This setup was kept consistent across all the libraries.
\subsubsection{GPU:} 

\begin{table}[!h]
  \caption{Test matrices and relevant attributes} 
  
  \label{tab:freq}
  \begin{tabular}{crrl}
    \toprule
        \thead{Matrix} & \thead{Dimension} & \thead{NNZ} & \thead{Name}\\ 
    \midrule
        A & 25,503 & 1.55e+04 & \verb|bcsstm37|\\ 
        B & 46,772 & 4.68e+04 & \verb|bcsstm39|\\ 
        C & 25,187 & 1.93e+05 & \verb|mult_dcop_01|\\ 
        D & 131,072 & 7.86e+05 & \verb|delaunay_n17| \\ 
        E & 41,092 & 1.68e+06 & \verb|av41092|\\ 
        F & 32,1671 & 1.83e+06 & \verb|ASIC320ks|\\ 

  \bottomrule
\end{tabular}
\end{table}
\Cref{fig:nnz_gpu} shows the SpMV speedups achieved by pyGinkgo, PyTorch, TensorFlow, and CuPy on an NVIDIA A100 GPU, relative to the baseline performance of SciPy running on a single CPU core. The matrices are ordered by increasing non-zero count. Among all libraries, pyGinkgo consistently outperforms the alternatives, exhibiting near-linear scaling with respect to the number of non-zeroes (NNZ).

For the evaluated matrices, pyGinkgo reaches a peak performance of approximately 150 GFLOP/s, followed by PyTorch at 110 GFLOP/s, CuPy at 85 GFLOP/s, and TensorFlow at 50 GFLOP/s. While PyTorch achieves relatively high peak performance, it is still approximately twice as slow as pyGinkgo across most test cases. CuPy is generally 3–4 times slower, and TensorFlow exhibits the largest performance gap, being 2–14 times slower than pyGinkgo, depending on the matrix.

\Cref{fig:bar_gpu} illustrates the speedup of selected matrices spanning several orders of magnitude in terms of NNZ. The relevant attributes of these representative matrices are listed in \Cref{tab:freq}. Across all libraries, a clear trend of increasing speedup with increasing NNZ can be observed. 

With pyGinkgo, we are able to run SpMV computations not only on NVIDIA GPUs but also on AMD GPUs, enabling a broader performance analysis across architectures. The performance comparison between NVIDIA A100 and AMD Instinct MI100 accelerators for SpMV is presented in \Cref{fig:spmv_perf}. The plot compares both CSR and COO matrix formats. We observe that performance on NVIDIA A100 is slightly better than that on AMD GPU, particularly for matrices with larger NNZ.

\subsubsection{CPU:}
On a single CPU thread, we observe that SciPy performs better than all the other libraries, although its performance does not scale well with increasing number of threads. pyGinkgo, on the other hand, scales well with increasing number of threads, its speedup relative to SciPy across multiple threads is shown in \Cref{fig:nnz_cpu}.  pyGinkgo's execution time for the SpMV kernel proves to be at least 7-35 times faster than SciPy for matrices with higher NNZ as we scale to 32 threads. Compared to PyTorch and TensorFlow, pyGinkgo is 10-60 and 30-90 times faster, respectively. As also shown for GPU earlier, the speedups across various libraries for the selected matrices is presented in \Cref{fig:bar_cpu}. We observe that it is highly beneficial to use pyGinkgo for large matrices (D and F). For the matrix E, the speedup drops across all libraries because of its higher density. Comparing \Cref{fig:bar_gpu} and \Cref{fig:bar_cpu}, we observe that it is more efficient to use CPU instead of GPU for matrices with low NNZ (A and B).

\subsection{Solvers}
To evaluate pyGinkgo's performance for solvers, we considered linear systems of the form $Ax = b$, where $A$ is a sparse matrix from the dataset, $b$ is initialized as a vector of ones, and the initial guess $x$ is a vector of zeros. This setup was kept consistent across CuPy, SciPy and pyGinkgo. We conducted a similar evaluation for solver kernels as we did for SpMV, using a benchmark set of 40 sparse matrices. 





\begin{figure*}[!h]
    \centering
    \begin{subfigure}[b]{0.30\textwidth}
        \centering
        \includegraphics[width=\textwidth]{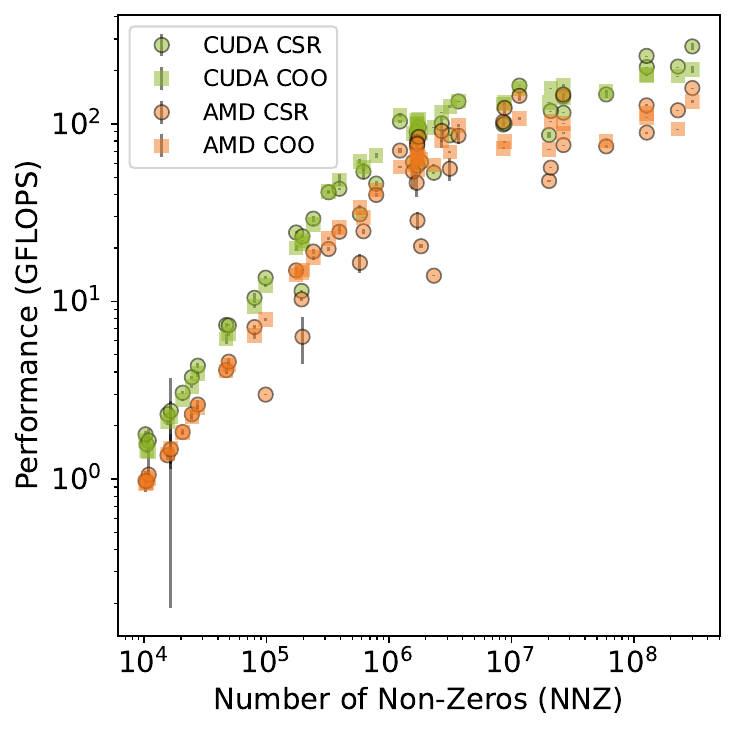}
        \caption{\small pyGinkgo's SpMV performance with number of non-zeros}
        \label{fig:spmv_perf}
    \end{subfigure}
    \hfill
    \begin{subfigure}[b]{0.30\textwidth}
        \centering
        \includegraphics[width=\textwidth]{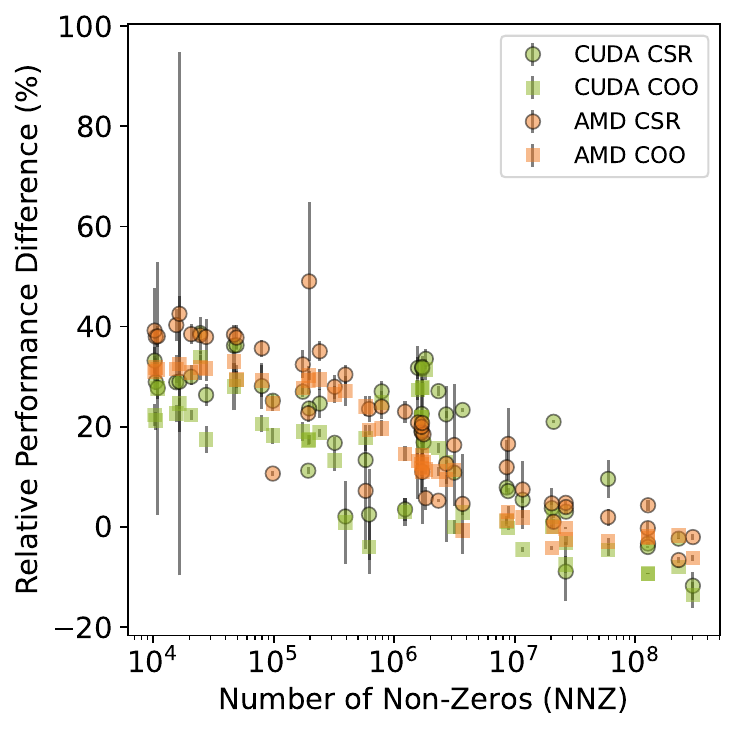}
        \caption{\small Relative Performance difference of pyGinkgo versus Ginkgo }
        \label{fig:perf_overhead}
    \end{subfigure}
    \hfill
    \begin{subfigure}[b]{0.30\textwidth}
        \centering
        \includegraphics[width=\textwidth]{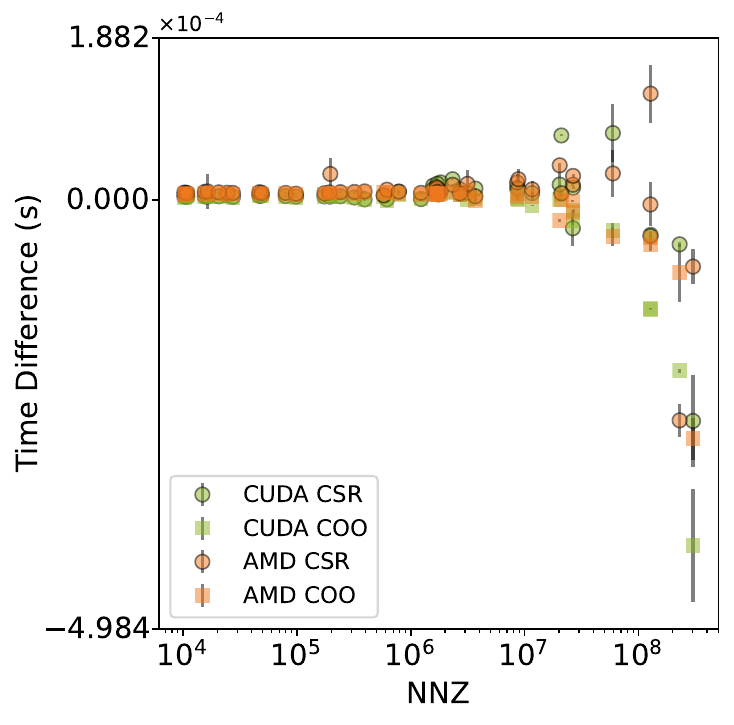}
        \caption{\small Time difference of pyGinkgo compared to Ginkgo in seconds }
        \label{fig:time_overhead}
    \end{subfigure}
    \caption{Performance difference and time difference analysis relative to native Ginkgo implementation for SpMV on NVIDIA A100 and AMD Instinct MI100 accelerators for CSR and COO matrix formats}
    \label{fig:perf_comparison}
\end{figure*}

\subsubsection{GPU:}
 While PyTorch and TensorFlow offer little support for sparse linear algebra, primarily supporting direct solvers, they do not provide implementations of iterative solvers. Consequently, our comparison in this section focuses on CuPy and pyGinkgo.

 CuPy supports a limited set of iterative Krylov subspace solvers, specifically the Conjugate Gradient (CG), the conjugate gradient squared method (CGS), generalized minimal residual (GMRES), least squares with QR factorization (LSQR), least squares minimization of residuals (LSMR) and the Minimum Residual (MINRES) methods. Since there is no native support for preconditioners on CuPy's side, this analysis was done without a preconditioner. It is worth noting, however, that pyGinkgo exposes a rich set of both iterative solvers and preconditioners. Our benchmark experiments were done for CG, CGS and GMRES for both CuPy and pyGinkgo, performing 1000 solver iterations for each case. Many of the SuiteSparse matrices we tested were ill-conditioned, and as a result, several of them did not converge, especially in the absence of a preconditioner. Therefore, it was more meaningful to evaluate performance based on time per iteration rather than total time to convergence. However, time per iteration alone does not reflect the total time required to reach convergence.

pyGinkgo's speedup relative to CuPy for all three solvers is presented in \Cref{fig:solver_nvidia}. CGS consistently achieves the highest speedup, particularly for matrices with fewer nonzeros, reaching up to 4 times than that of CuPy. CG offers moderate speedup of around 2.5 times across a wide range of NNZ values. We observe that the speedup for pyGinkgo decreases noticeably with the number of nonzeros, This result suggests possible overheads or optimization thresholds on CuPy’s side. 

Although pyGinkgo generally outperforms CuPy for CG and CGS solvers, CuPy shows slightly better performance for GMRES. Here, the restart factor for GMRES was chosen to be $30$.

The GMRES implementations in CuPy and Ginkgo differ in several important ways. First, CuPy solves the least-squares problem associated with the Hessenberg matrix on the CPU, whereas Ginkgo performs the entire solution process on the GPU. Although GPU-based computation generally offers higher throughput, solving very small Hessenberg systems may actually be faster on the CPU due to low parallelization from dependence requirement in the triangular solver. Second, Ginkgo updates the Hessenberg matrix by using Givens rotations, while CuPy employs an orthonormal projection approach. Third, Ginkgo checks the residual norm after every update to the Hessenberg matrix, while CuPy only performs the check after the complete Hessenberg matrix is constructed, resulting in $restart-1$ additional checks in Ginkgo. Additionally, Ginkgo reuses the computed Givens rotations to efficiently update the residual norm. These differences in strategy, particularly the more frequent residual checks and the full GPU implementation may contribute to CuPy's slightly faster performance compared to pyGinkgo in GMRES especially when using a fixed number of iterations rather than convergence-based stopping. 

\subsubsection{CPU:}
We also executed these benchmarks on CPU to compare it with SciPy. pyGinkgo is around 3 - 8 times faster than Scipy in the case of CG for the same systems of matrices that we considered for CuPy. We obtained similar results for CGS and GMRES as well. 

\subsection{pyGinkgo versus Ginkgo}
\label{section:pgko_vs_gko}

In addition to benchmarking against Python-based libraries, we also benchmarked pyGinkgo against Ginkgo's native SpMV to assess the overhead of the Python bindings. Results for the GPU backends are presented in \Cref{fig:perf_overhead} and \Cref{fig:time_overhead}. 
We evaluated the performance and the corresponding time difference of pyGinkgo relative to the native Ginkgo implementation using the following metrics.

The relative performance difference is defined as:
\begin{equation*}
    P_{\text{overhead}} = \frac{P_{\text{Ginkgo}} - P_{\text{pyGinkgo}}}{P_{\text{Ginkgo}}} \times 100
\end{equation*}
where $P_{\text{overhead}}$ represents the relative performance difference between pyGinkgo and Ginkgo, $P_{\text{Ginkgo}}$ is the performance of Ginkgo in GFLOPS, and $P_{\text{pyGinkgo}}$ is the performance of pyGinkgo in GFLOPS for the corresponding matrix format and hardware backend.

The time difference is computed as:
\begin{equation*}
    T_{\text{overhead}} = T_{\text{pyGinkgo}} - T_{\text{Ginkgo}}
\end{equation*}
where $T_{\text{overhead}}$ is the additional execution time in seconds introduced by pyGinkgo, with $T_{\text{pyGinkgo}}$ and $T_{\text{Ginkgo}}$ denoting the SpMV execution times of pyGinkgo and Ginkgo, respectively.

\subsubsection{NVIDIA GPU:} Results indicate that for NVIDIA GPUs, the performance overhead of pyGinkgo is around $30\%$ for matrices with lower NNZ. It substantially decreases as number of non-zeros increase. For both matrix formats, overheads drop from approximately $25 - 35\%$ at lower NNZ to below $10\%$ for larger NNZ (NNZ $> 10^7$) for most cases. However, the time difference, as presented in \Cref{fig:time_overhead}, mostly remains in the order of $10^{-7} - 10^{-5}$ seconds for all test matrices, which is negligible and unlikely to be of practical concern, especially for matrices with low NNZ. In the cases for matrices with larger NNZ, the time difference can sometimes be below zero due to variability from system noise, which prevents perfectly consistent timing results. Although we followed the same benchmarking procedure on both the Python and C++ sides, the timing mechanisms differ: we use \texttt{steady\_clock} from the C++ standard library and the time module in Python, both after explicit GPU synchronization. These differences in timer implementation and the effects of synchronization introduce some measurement uncertainty. Nevertheless, as shown in \Cref{fig:perf_overhead} and \Cref{fig:time_overhead}, the results consistently indicate that pyGinkgo adds negligible overhead for large-scale problems on NVIDIA GPUs.
\subsubsection{AMD GPU:} On AMD GPUs, the performance overhead introduced by pyGinkgo is slightly higher as compared to NVIDIA GPUs, particularly for the CSR format. At smaller matrix sizes, the percentage overhead can exceed 40\% for a few cases, with notable fluctuations across both CSR and COO formats. The absolute time difference remains relatively low - typically within the range of $10^{-6} - 10^{-4}$ seconds for all test matrices. Similar to the observation on NVIDIA GPUs, the time difference below zero indicates that the machine noise, differences in timer implementation and the effects of synchronization lead to that difference. These results suggest that pyGinkgo incurs low overhead on AMD hardware as well, especially for matrices with larger NNZ.


\section{Conclusion and Outlook}

In this work, we brought pyGinkgo, a Python wrapper for Ginkgo, providing high-performance sparse matrix operations and linear solvers on CPUs and GPUs. Our benchmarks show that pyGinkgo significantly outperforms existing Python libraries such as SciPy, CuPy, TensorFlow, and PyTorch for sparse matrix-vector multiplication. For iterative solvers, pyGinkgo demonstrates performance comparable to that of CuPy, further validating its efficiency for workloads with sparse linear solvers. This makes pyGinkgo a strong candidate for serving as the computational backbone of sparse scientific computing and machine learning workloads.

Until now, Python users have had limited access to optimized sparse operations across modern architectures, often sacrificing performance for simplicity. pyGinkgo changes this by combining Ginkgo’s state-of-the-art, hardware-accelerated backend with a lightweight, NumPy-compatible, and Pythonic interface. Users are now able to use SpMV kernels, advanced iterative solvers, and preconditioners without writing C++ code. We are actively extending the library with new algorithmic capabilities, and we are also implementing iterative methods like the Rayleigh-Ritz algorithm entirely on the Python side to support advanced eigensolvers. Our framework’s flexibility also allows users to implement custom algorithms easily by combining existing linear algebra operations accessible via pyGinkgo.

On the Ginkgo side, future work includes the integration of a convolution kernel, which would allow Ginkgo and pyGinkgo to support key operations required in image processing and convolutional neural networks. With continuous developments, pyGinkgo interface will become even more Pythonic and user-friendly, enhancing usability for the machine learning and scientific computing communities.
In essence, pyGinkgo makes high-performance sparse computing accessible to the Python community. It opens the door for scientists, engineers, and machine learning practitioners to write high-level Python code while still benefiting from the raw speed and scalability of Ginkgo.

\begin{acks}
This work was performed on the HoreKa supercomputer and the NHR@KIT Future Technologies Partition testbed, funded by the Ministry of Science, Research and the Arts Baden-Württemberg and by the Federal Ministry of Research, Technology and Space. This work is supported by the Helmholtz AI platform grant. The authors would like to thank the Federal Ministry of Education and Research and the state governments (www.nhr-verein.de/unsere-partner) for supporting this
work/project as part of the joint funding of National High Performance Computing
(NHR). We are also grateful to Dr. Yoshifumi Nakamura and Dr. Hitoshi Murai from RIKEN for their meaningful discussions and valuable insights. The authors acknowledge the use of AI-based tools for language editing and grammar improvement during the preparation of this manuscript.
\end{acks}

\bibliographystyle{ACM-Reference-Format}
\bibliography{references}

\appendix

\end{document}